\documentclass[aip,amsmath,amssymb,reprint]{revtex4-1}

\usepackage{graphicx}
\usepackage{dcolumn}
\usepackage{bm}
\usepackage[mathlines]{lineno}

\usepackage[x11names]{xcolor}   
\definecolor{DarkGreen}{rgb}{0.0,0.5,0.0}
\usepackage[bookmarksopen,bookmarksnumbered,colorlinks,citecolor=red,urlcolor=blue,filecolor=green]{hyperref}
\usepackage{tabularray}
\definecolor{DarkGreen}{rgb}{0,0.7,0}

\usepackage[utf8]{inputenc}
\usepackage[T1]{fontenc}
\usepackage{mathptmx}
\usepackage{etoolbox}
            
\makeatletter
\def\@email#1#2{%
 \endgroup
 \patchcmd{\titleblock@produce}
  {\frontmatter@RRAPformat}
  {\frontmatter@RRAPformat{\produce@RRAP{*#1\href{mailto:#2}{#2}}}\frontmatter@RRAPformat}
  {}{}
}%
\makeatother
\graphicspath{{figures/v4/}}

\begin{document}
\preprint{AIP/123-QED}

\title{Kinetically Controlled Condensation Boundary Governing Indium Incorporation in InGaN Metal--Organic Vapor Phase Epitaxy} 

\author{Qihui Lin}
	\affiliation{State Key Laboratory of Artificial Microstructure and Mesoscopic Physics, School of Physics, Peking University, Beijing 100871, P. R. China}

\author{Junlin Wu}
	\affiliation{State Key Laboratory of Artificial Microstructure and Mesoscopic Physics, School of Physics, Peking University, Beijing 100871, P. R. China}
    
\author{Erqi Xu}
	\affiliation{State Key Laboratory of Artificial Microstructure and Mesoscopic Physics, School of Physics, Peking University, Beijing 100871, P. R. China}

\author{Jiaqing Yue}
	\affiliation{State Key Laboratory of Artificial Microstructure and Mesoscopic Physics, School of Physics, Peking University, Beijing 100871, P. R. China}

\author{Jiale Wang}
	\affiliation{State Key Laboratory of Artificial Microstructure and Mesoscopic Physics, School of Physics, Peking University, Beijing 100871, P. R. China}

    \author{Zihao Xu}
	\affiliation{State Key Laboratory of Artificial Microstructure and Mesoscopic Physics, School of Physics, Peking University, Beijing 100871, P. R. China} 

\author{Haixin Qi}
	\affiliation{State Key Laboratory of Artificial Microstructure and Mesoscopic Physics, School of Physics, Peking University, Beijing 100871, P. R. China} 
    
\author{Liyi Luo}
	\affiliation{State Key Laboratory of Artificial Microstructure and Mesoscopic Physics, School of Physics, Peking University, Beijing 100871, P. R. China} 

\author{Haitao Wang}
	\affiliation{Institute of Materials and Systems for Sustainability, Nagoya University, Nagoya, Japan}
    
\author{Jia Wang}
	\affiliation{Institute for Advanced Research, Nagoya University, Nagoya, Japan}
    \affiliation{Institute of Materials and Systems for Sustainability, Nagoya University, Nagoya, Japan}

\author{Hiroshi Amano}
	\affiliation{Institute for Advanced Research, Nagoya University, Nagoya, Japan}
    \affiliation{Institute of Materials and Systems for Sustainability, Nagoya University, Nagoya, Japan}

\author{Bo Shen}
	\affiliation{State Key Laboratory of Artificial Microstructure and Mesoscopic Physics, School of Physics, Peking University, Beijing 100871, P. R. China}
    
\author{Guangxu Ju}
    \email[correspondence to: ]{gxju@pku.edu.cn}
	\affiliation{State Key Laboratory of Artificial Microstructure and Mesoscopic Physics, School of Physics, Peking University, Beijing 100871, P. R. China}


\date{\today}

\begin{abstract}

We combine \textit{in situ} synchrotron X-ray crystal truncation rod measurements with a binary Burton--Cabrera--Frank model to quantify indium incorporation during In$_x$Ga$_{1-x}$N growth by metal–organic vapor phase epitaxy (MOVPE) on GaN(0001). By distinguishing In adatoms from condensed droplets and incorporating coupled Ga–In incorporation kinetics, the model captures the intrinsically nonlinear dependence of indium composition $x_{\mathrm{In}}$ on flux and temperature. The critical In coverage corresponding to  the maximum attainable In composition at a given temperature is determined by a kinetic balance between In adatom supply and incorporation capacity, defining a kinetically controlled condensation boundary that shifts with temperature and Ga flux. The model quantitatively predicts this boundary, consistent with independent measurements, and provides a framework for optimizing high-In-content InGaN growth while avoiding droplet formation.

\end{abstract}


\maketitle 


Indium-containing III-nitride alloys, particularly InGaN, are central to modern optoelectronic devices, enabling efficient blue and green light emission and offering a pathway toward monolithic RGB light sources. \cite{ amano2016development} Despite their technological importance, precise control of indium incorporation during metal--organic vapor phase epitaxy (MOVPE) remains a challenge.

A key difficulty arises from the nontrivial relationship between indium supply and incorporation. While increasing the indium flux is expected to enhance incorporation, experiments show that above a critical flux, metallic indium droplets form on the surface, reducing the indium content in the growing film. \cite{ou1998elucidation,bedair1997growth} \textit{In situ} X-ray fluorescence studies further demonstrate that, below the kinetic condensation boundary, the surface indium coverage follows a Langmuir-type behavior and saturates at a sub-monolayer level ($\sim 0.25$ ML).\cite{2006_Jiang_APL89_161915} However, such adsorption-based descriptions do not explain how surface indium is incorporated during growth.

Recent work has shown that indium incorporation is strongly influenced by the local chemical environment, with indium atoms preferentially incorporated when adjacent to Ga atoms.\cite{hu2022understanding} This indicates that incorporation is governed not only by surface coverage but also by local atomic configurations, introducing an intrinsic coupling between surface composition and incorporation pathways.

The Burton–Cabrera–Frank (BCF) framework describes adatom diffusion and incorporation kinetics on crystal surfaces\cite{1951_Burton_PhilTransRS243_29,ghez1988kinetics,jeong1999steps,krug2005introduction,2015_Woodruff_PhilTransA373_20140230} and has been extended to account for anisotropic step kinetics on GaN(0001).\cite{2022_Ju_PRB105_054312,ju2024burton} However, a quantitative connection between surface indium coverage and incorporation during growth remains lacking. Thermodynamic models have described vapor--solid equilibrium and droplet formation in InGaN growth,\cite{koukitu1997thermodynamic} but do not explicitly establish this connection with the incorporation process.

In this Letter, we combine \textit{in situ} synchrotron X-ray crystal truncation rod (CTR) measurements with a binary (two-species) BCF model to quantify indium incorporation during InGaN MOVPE on GaN(0001). By distinguishing In adatoms from metallic In droplets and incorporating coupled Ga–In incorporation, the model links adsorption-limited surface coverage to incorporation kinetics. We show that $x_{\mathrm{In}}$ decreases once the TMIn flux exceeds a critical value, coinciding with droplet formation. This critical condition shifts with temperature and Ga flux because it is governed by a kinetic balance between In adatom supply and incorporation capacity. The resulting kinetic condensation boundary can be quantitatively predicted, providing a framework for identifying optimal growth conditions for high-In-content InGaN.

InGaN epitaxial growth was carried out at beamline 12-ID-D of the Advanced Photon Source using a vertical-flow MOVPE reactor integrated with a diffractometer, enabling \textit{in situ} X-ray measurements during growth.\cite{2017_Ju_RSI88_035113} This system has previously been used for coherent X-ray studies of GaN surface structure and growth dynamics.\cite{2021_Ju_NatCommun12_1721, 2019_Ju_NatPhys15_589} To penetrate the 2-mm-thick quartz walls of the reactor, an X-ray energy of 28.8 keV was employed.

All growth experiments were performed on a single GaN(0001) substrate with a 0.54$^\circ$ off-cut.\cite{wangdetermination} Prior to deposition, a thin GaN regrowth layer ($\sim$ 100 nm) was grown \textit{in situ} at high temperature to obtain a clean and well-ordered surface. Triethylgallium (TEGa), trimethylindium (TMIn), and ammonia (NH$_3$) were used as precursors, with nitrogen as the carrier gas at a total pressure of 200 Torr. The NH$_3$ flow was fixed at $4.91\times10^{-2}$ mol/min (1.1 slpm) throughout all experiments.

InGaN layers were grown under three sets of conditions, in which only one growth parameter was varied at a time: (i) TMIn flow at fixed temperature and TEGa flow, (ii) TEGa flow at fixed temperature and TMIn flow, and (iii) temperature at fixed TMIn and TEGa flows. The TMIn flow was varied from 0.0457 to 0.914~$\mu$mol/min, the TEGa flow from 0.329 to 1.32~$\mu$mol/min and the temperature from 939 to 1049~K. Representative growth conditions (Cond.) are summarized in Table~\ref{xIn_fit_table_thiswork}, while the complete dataset and precursor-flow conversion are provided in Table~S3 and Sec.~I of the Supplementary Material.\cite{2026_Lin_APL_supplemental} The corresponding V/III ratio ranged from 2.20$\times10^4$ to 6.97$\times10^4$. The InGaN layer thickness was typically in the range of 3.17–5.88~nm (12.2-22.6~ML), which was below the critical thickness for strain relaxation.\cite{ju2017role} 

After each growth step, the sample was characterized \textit{in situ} by X-ray CTR scattering using symmetric $(0002)$ scans.\cite{ju2014situ,ju2011x} Prior to measurement, the TMIn flow was reduced to a fixed low value (0.0914~$\mu$mol/min) for all samples and maintained during the measurement to establish a surface adsorption--desorption equilibrium, thereby suppressing net In loss from the InGaN layer. Following characterization, the InGaN layer was removed by a brief high-temperature anneal in H$_2$, restoring the surface to bare GaN for subsequent growth cycles. Cond.~4 in Table~S1\cite{2026_Lin_APL_supplemental} was intentionally repeated to verify reproducibility of the growth–anneal–regrowth procedure. 

The indium composition $x_{\mathrm{In}}$ and growth rate $G$ are extracted from symmetric $(0002)$ CTR profiles using a kinematic scattering model for a strained InGaN layer on vicinal GaN(0001).\cite{2026_PKU_PRB} The model includes coherent scattering from the substrate, epilayer, and vicinal surface, together with full elastic lattice distortion,\cite{2006_romanov_JAP100_023522} including Nagai tilt and shear-induced triclinic deformation.\cite{1974_Nagai_JAP45_3789,2013_Krysko_JAP114_113512,2021_Liao_JAP129_025105,2026_PKU_PRB} Because the CTR line shape separates thickness and composition sensitivities through interference fringes and lattice distortion, respectively, $x_{\mathrm{In}}$ and film thickness can be determined with reduced parameter covariance.\cite{2005_Schleputz_ACA61_418} The growth rate $G$ is then obtained from the fitted thickness and deposition time. The fitted roughness ranges from 0.16 to 0.35~nm, consistent with sub-monolayer step-flow growth and supporting the applicability of the BCF framework. Further details are provided in the Supplementary Material.\cite{2026_Lin_APL_supplemental}

Consistent with the \textit{ex situ} study,\cite{ou1998elucidation} where indium droplets emerge as $x_{\mathrm{In}}$ begins to decrease, the onset of droplet formation is identified with the maximum of $x_{\mathrm{In}}$. We thus define the condensation boundary as this kinetic transition point, separating incorporation-limited and droplet-dominated regimes.

We describe surface occupation using a Langmuir-type adsorption model, in which the total indium coverage is limited by the finite number of surface sites.\cite{2006_Jiang_APL89_161915} Both mobile In adatoms ($\theta_{\mathrm{In}}$) and condensed In droplets ($\theta_{\mathrm{In\_d}}$) contribute to the occupation of these sites, such that
\begin{equation}
\theta_{\mathrm{In}}^{\mathrm{tot}} = \theta_{\mathrm{In}} + \theta_{\mathrm{In\_d}},
\label{totalcoverage}
\end{equation}
where $\theta_{\mathrm{In}}^{\mathrm{tot}} \le \theta^{\mathrm{sat}}_{\mathrm{In}}$ and  $\theta^{\mathrm{sat}}_{\mathrm{In}}= 0.25$ corresponds to the reported saturation coverage for In adsorption on GaN(0001).\cite{2006_Jiang_APL89_161915}

\begin{figure}[t]
    \centering
    \includegraphics[width=1\linewidth]{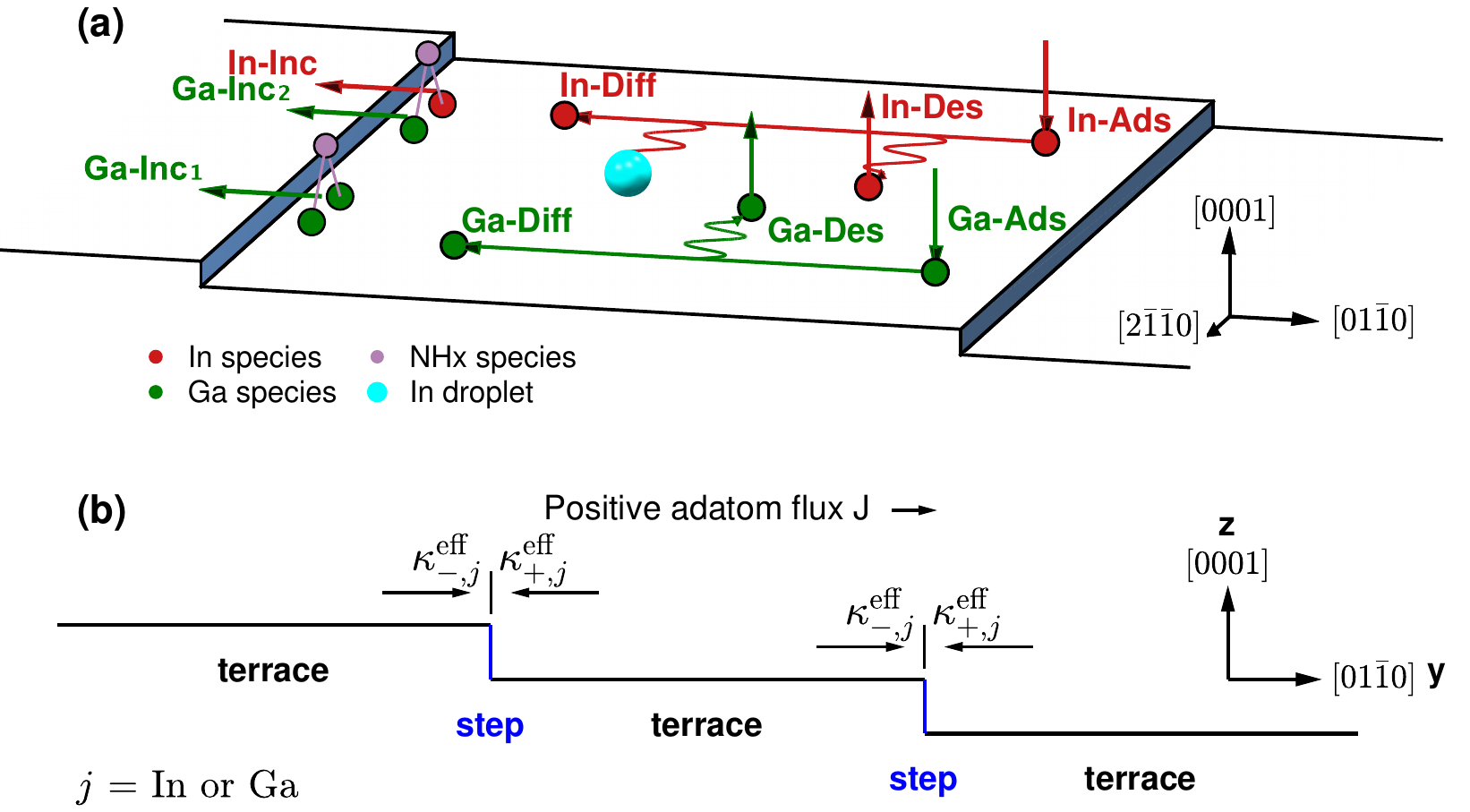}
    \caption{(a) Schematic of InGaN growth on a vicinal surface. In and Ga adatoms undergo adsorption, diffusion, desorption, and incorporation; Ga incorporates independently or with In, whereas In incorporation is predominantly Ga-assisted in the present growth regime. Excess In forms droplets at high flux. (b) Binary BCF model geometry showing terraces, steps, and attachment from the upper ($\kappa_{-,j}^\mathrm{eff}$) and lower ($\kappa_{+,j}^\mathrm{eff}$) terraces. }
    \label{fig:placeholder}
\end{figure}

While this Langmuir description captures site occupation, it does not describe step-edge incorporation or distinguish different incorporation pathways. We therefore formulate a binary BCF model for the coupled dynamics of Ga and In adatoms on the same terrace, with metallic In droplets treated as a distinct surface species. 
Condensed In droplets are assumed to be kinetically decoupled from step-edge incorporation and therefore do not contribute directly to the incorporation flux. The incorporation pathways are illustrated in Fig.~\ref{fig:placeholder}(a). Ga adatoms can incorporate either independently or through pairing with In adatoms (Ga--Ga or Ga--In pairs), whereas In incorporation requires neighboring Ga atoms. Thus, Ga adatoms do not directly limit Langmuir-type In adsorption, but they control the availability of Ga-assisted In incorporation pathways. In contrast, In adatoms and In droplets partition the same finite In surface reservoir, constrained by Eq.~(\ref{totalcoverage}). This coupling links adsorption-limited coverage, droplet formation, and step-edge incorporation kinetics within a unified BCF framework.

Based on these considerations, the surface densities per unit area, $\rho_j$, evolve according to the following continuity equations, where Ga and In denote gallium and indium adatoms, respectively, and $\mathrm{In\_d}$ denotes indium in the droplet phase.
\begin{equation}
\begin{aligned}
\frac{\partial \rho_{\mathrm{Ga}}}{\partial t}
&= D_{\mathrm{Ga}} \nabla_y^2 \rho_{\mathrm{Ga}}
- \frac{\rho_{\mathrm{Ga}}}{\tau_{\mathrm{Ga}}}
+ F_{\mathrm{Ga}}, \\[6pt]
\frac{\partial \rho_{\mathrm{In}}}{\partial t}
&= D_{\mathrm{In}} \nabla_y^2 \rho_{\mathrm{In}}
- \frac{\rho_{\mathrm{In}}}{\tau_{\mathrm{In}}}
+ F_{\mathrm{In}}\!\left(1 - \frac{\rho_{\mathrm{In}} + \rho_{\mathrm{In\_d}}}{\rho_{\mathrm{In}}^{\mathrm{sat}}}\right), \\[6pt]
\frac{\partial \rho_{\mathrm{In\_d}}}{\partial t}
&= - \frac{\rho_{\mathrm{In\_d}}}{\tau_{\mathrm{In\_d}}}
+ F_{\mathrm{In}}\!\left(1 - \frac{\rho_{\mathrm{In}} + \rho_{\mathrm{In\_d}}}{\rho_{\mathrm{In}}^{\mathrm{sat}}}\right).
\end{aligned}
\label{BCF_model}
\end{equation}
Here, $\rho_j^{\mathrm{sat}}$, $D_j$, and $F_j$ denote the saturation surface density, diffusivity,\cite{zywietz1998adatom} and deposition flux, respectively. The parameter $\tau_j$ denotes the effective lifetime before removal from the adatom population. For Ga adatoms, which are removed primarily by desorption, $\tau_{\mathrm{Ga}}$ is treated as a temperature-dependent parameter that remains fixed at a given growth temperature.\cite{koleske2014connection} In contrast, In adatoms, with lifetime $\tau_{\mathrm{In}}$, can either desorb or convert into droplets, while $\tau_{\mathrm{In\_d}}$ characterizes the residence time of In in the droplet phase.

The boundary conditions for the flux at the step edges terminating the terrace are written as:
\begin{equation}
\label{Jj}
J_j^{\pm} =
\begin{cases}
- D_j \,\nabla_y \rho_j^{\pm}= \pm \kappa_{\mp,j}^{{\mathrm{eff}}} \bigl(\rho_j^{\pm} - \rho_{j}^{\mathrm{eq}}\bigr),& j = \mathrm{Ga~or~In},\,
\\[8pt]
0,
& j = \mathrm{In\_d},
\end{cases}
\end{equation}
where $J_{j}$ and $\rho_{j}^{\mathrm{eq}}$ denote the surface flux and equilibrium adatom densities of species $j$, respectively. The superscripts $+$ and $-$ indicate evaluation at the lower and upper terrace boundaries, respectively, as illustrated in Fig.~\ref{fig:placeholder}(b). Here, $w$ is the terrace width, $y=0$ denotes the terrace center. 

\begin{figure}
    \centering
    \includegraphics[width=1\linewidth]{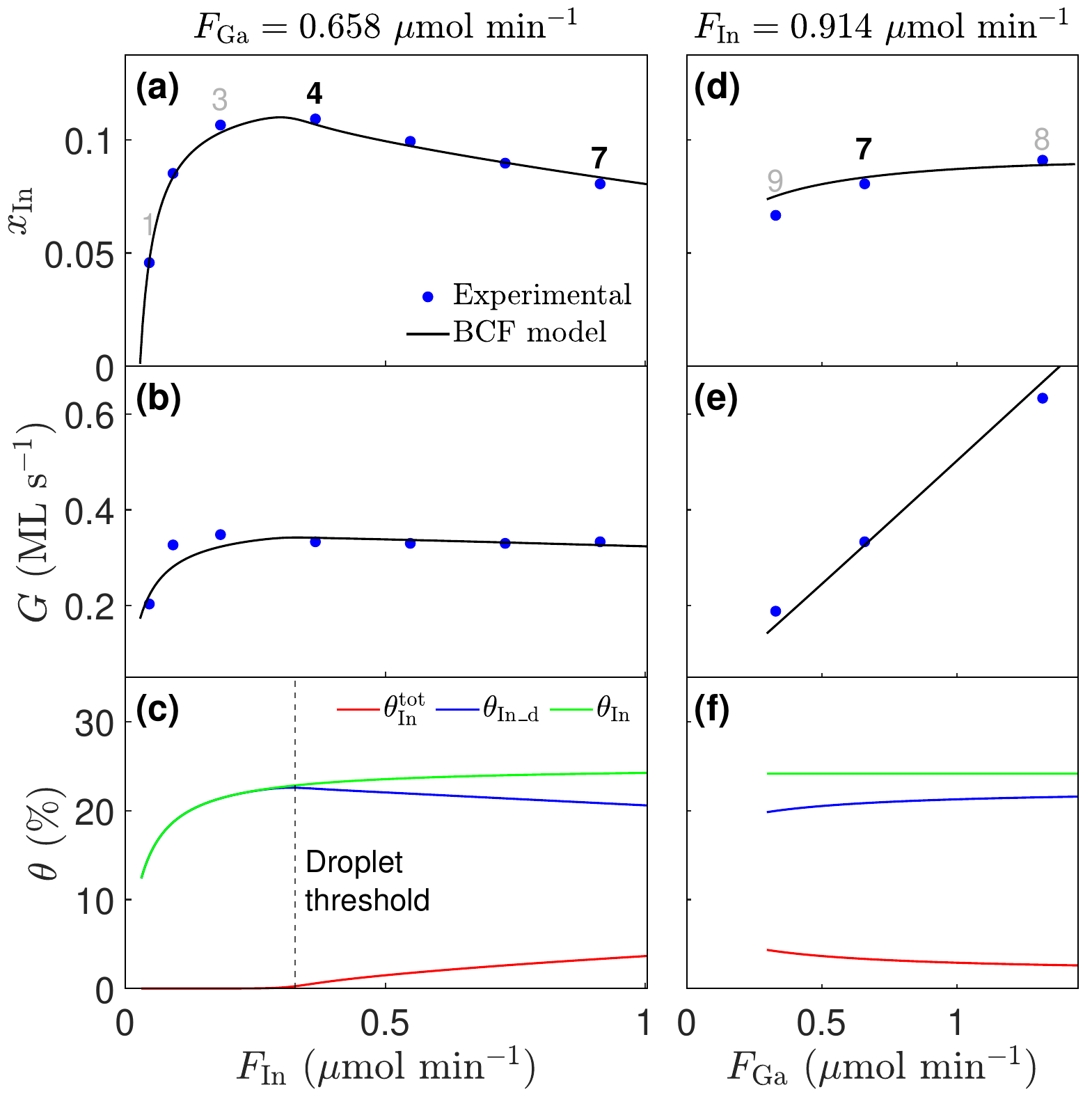}
    \caption{In composition $x_{\mathrm{In}}$, growth rate $G$, and surface coverages at $T=1005$~K. (a–c) vs $F_{\mathrm{In}}$ at fixed $F_{\mathrm{Ga}}$; (d–f) vs $F_{\mathrm{Ga}}$ at fixed $F_{\mathrm{In}}$. Conds.~1, 3, 4, 7, 8, and 9 correspond to Table~\ref{xIn_fit_table_thiswork}. Cond.~7 is shared by the constant-$F_{\mathrm{Ga}}$ and constant-$F_{\mathrm{In}}$ series.}   
    \label{fit_1005}
\end{figure}

The effective attachment coefficient $\kappa_{\mp,j}^{{\mathrm{eff}}} $ combines the intrinsic step-edge attachment kinetics $\kappa_{\mp}^{j}$ with the incorporation pathway factor $\eta_j$, which accounts for the influence of the local surface composition on incorporation. For In incorporation, $\eta_{\mathrm{In}} = K_1 \theta_{\mathrm{Ga}} \frac{\theta_{\mathrm{In}}^{\mathrm{tot}}-\theta_{\mathrm{In\_d}}}{\theta_{\mathrm{In}}^{\mathrm{tot}}}$ describes Ga-assisted incorporation of In adatoms. For Ga incorporation, $\eta_{\mathrm{Ga}} = K_2 \theta_{\mathrm{In}} + \left(1 + K_3 \frac{\theta_{\mathrm{In\_d}}}{\theta_{\mathrm{In}}^{\mathrm{tot}}}\right)$, which includes both cooperative Ga--In incorporation ($K_2 \theta_{\mathrm{In}}$) and a baseline Ga incorporation channel (unity term). 
The last term accounts for the continued incorporation of Ga through a Ga-only pathway as surface In becomes partitioned into droplets and is required to reproduce the weak dependence of the growth rate on TMIn flux observed experimentally [Fig.~\ref{fit_1005}(b)]. Together, these pathway factors describe the coupled incorporation of In and Ga during growth.\cite{keller1996growth,bedair1997growth,neugebauer2003adatom}

At quasi-steady state, the general solution for $\rho_j$ satisfying Eq.~(\ref{BCF_model}) with $\partial \rho_j/\partial t = 0$ is given by
\begin{equation}
\label{rho_j}
\rho_j = \rho_j^{\infty} +
\begin{cases}
C_{1,j}\cosh(k_j y) + C_{2,j}\sinh(k_j y),
& j = \mathrm{Ga~or~In}, \\[6pt]
0,
& j = \mathrm{In\_d},
\end{cases}
\end{equation}
where $\rho_j^{\infty}$ is the steady‑state uniform density on an infinite terrace, and $C_{1,j}$ and $C_{2,j}$ are determined by the boundary conditions, and $k_j$ is given in the Supplementary Material.\cite{2026_Lin_APL_supplemental} Defining $c_j \equiv \cosh(k_j w/2)$, $s_j \equiv \sinh(k_j w/2)$, and the dimensionless kinetic parameters $p_j \equiv \kappa_{+,j}^\mathrm{eff}/(k_j D_j)$ and $q_j \equiv \kappa_{-,j}^\mathrm{eff}/(k_j D_j)$, analytical expressions for $C_{1,j}$ and $C_{2,j}$ are obtained (see Supplementary Material\cite{2026_Lin_APL_supplemental}).

The quasi-steady-state step velocity $v_j$ is determined by the net adatom flux arriving at the step edges,
\begin{equation}
\label{vj}
v_j = \frac{J_j^+ - J_j^-}{\rho_j^{\mathrm{sat}}}
= -2 D_j \frac{C_{1,j}\, k_j\, s_j}{\rho_j^{\mathrm{sat}}}, ~~~~~j = \mathrm{Ga~or~In}.
\end{equation}

The net growth rate $G$, expressed in monolayers per second (ML/s), is given by
\begin{equation}
\label{G}
\begin{aligned}
G &= \frac{v_{\mathrm{In}} + v_{\mathrm{Ga}}}{w} = -\frac{2}{w}\left(
D_{\mathrm{In}} \frac{C_{\mathrm{1,In}}\, k_{\mathrm{In}}\, s_{\mathrm{In}}}{\rho_{\mathrm{In}}^{\mathrm{sat}}}
+
D_{\mathrm{Ga}} \frac{C_{\mathrm{1,Ga}}\, k_{\mathrm{Ga}}\, s_{\mathrm{Ga}}}{\rho_{\mathrm{Ga}}^{\mathrm{sat}}}
\right).
\end{aligned}
\end{equation}

The indium composition is then obtained as
\begin{equation}
\label{xIn}
\begin{aligned}
x_{\mathrm{In}} &= \frac{v_{\mathrm{In}}}{v_{\mathrm{In}} + v_{\mathrm{Ga}}}
= \frac{
D_{\mathrm{In}} \frac{C_{\mathrm{1,In}}\, k_{\mathrm{In}}\, s_{\mathrm{In}}}{\rho_{\mathrm{In}}^{\mathrm{sat}}}
}{
D_{\mathrm{In}} \frac{C_{\mathrm{1,In}}\, k_{\mathrm{In}}\, s_{\mathrm{In}}}{\rho_{\mathrm{In}}^{\mathrm{sat}}}
+
D_{\mathrm{Ga}} \frac{C_{\mathrm{1,Ga}}\, k_{\mathrm{Ga}}\, s_{\mathrm{Ga}}}{\rho_{\mathrm{Ga}}^{\mathrm{sat}}}
}.
\end{aligned}
\end{equation}

In the present BCF model, the contributions of In adatoms and droplets follow Langmuir-type expressions with a common saturation denominator, reflecting competition for the same surface sites [see Supplementary Material,\cite{2026_Lin_APL_supplemental} Eq.~(S20)]. Accordingly, the total coverage can be recast as
\begin{equation}
\frac{1}{\theta_{\mathrm{In}}^{\mathrm{tot}}}
=
\frac{1}{\theta^{\mathrm{sat}}_{\mathrm{In}}}
+
\frac{\rho_0}{\tau_{\mathrm{In}}^{\mathrm{tot}}}
\frac{1}{F_{\mathrm{In}}},
\label{theta_linear}
\end{equation}
where $\rho_0$=$1.13 \times 10^{19}$~m$^{-2}$ is the density of lattice sites per unit area,\cite{2026_Lin_APL_supplemental} $\theta^{\mathrm{sat}}_{\mathrm{In}} = \rho_{\mathrm{In}}^{\mathrm{sat}}/\rho_0$, and $\tau_{\mathrm{In}}^{\mathrm{tot}}\equiv\tau_{\mathrm{In}}+\tau_{\mathrm{In\_d}}$ denotes an effective residence time that arises naturally from the additive contributions of adatom and droplet states. This quantity should be interpreted as an effective time scale, rather than a strictly additive microscopic lifetime. 

\begin{figure}[t]    
\centering    
\includegraphics[width=1\linewidth]{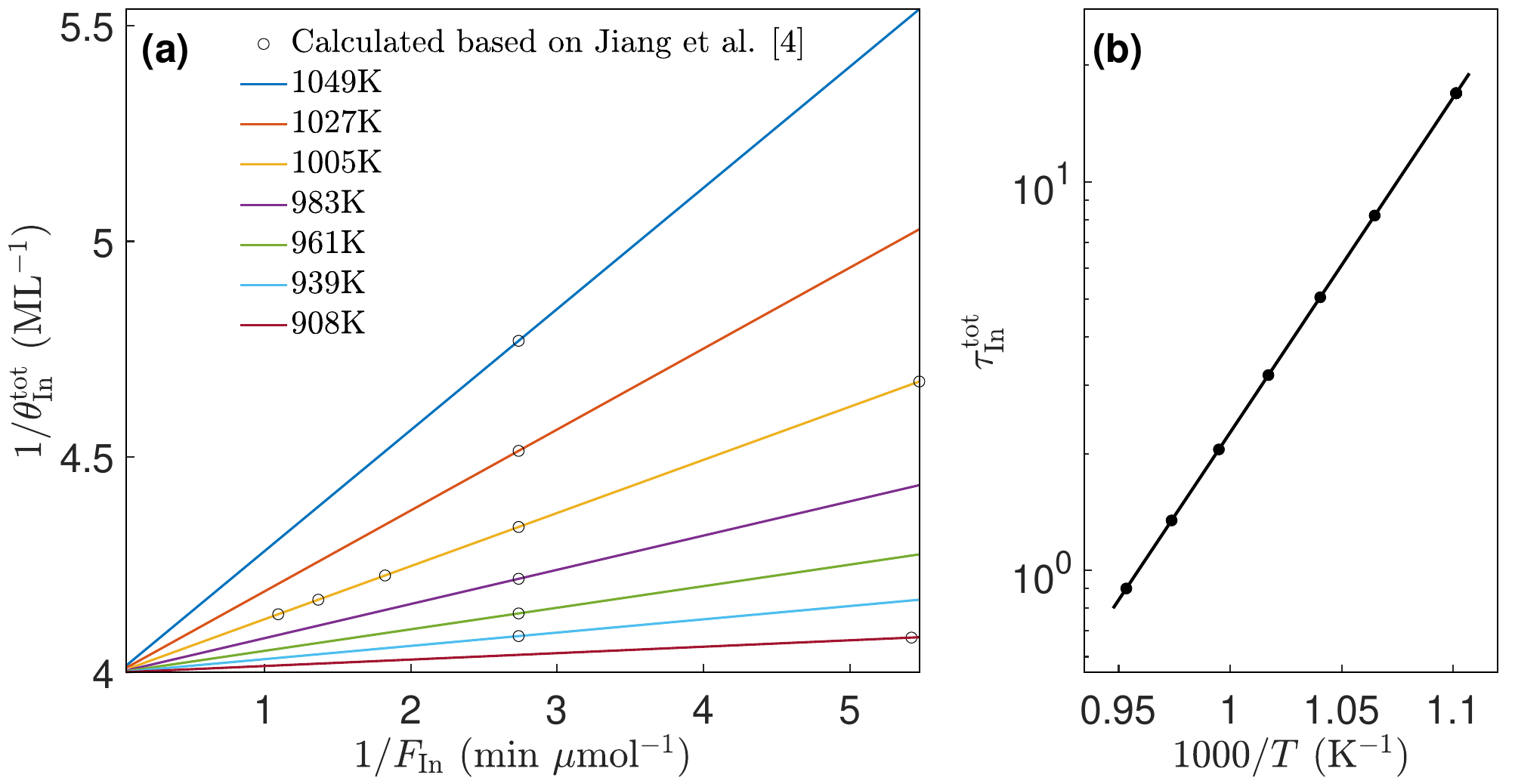}    
\caption{(a) $1/\theta_{\mathrm{In}}^{\mathrm{tot}}$ vs $1/F_{\mathrm{In}}$ at different temperatures. Symbols: calculated from the Langmuir model\cite{2006_Jiang_APL89_161915} (see Supplementary Material, Table~S1\cite{2026_Lin_APL_supplemental}); lines: present binary BCF model [Eq.~(\ref{theta_linear})]. (b) Temperature dependence of $\tau_{\mathrm{In}}^{\mathrm{tot}}$ extracted from the slopes in panel (a), establishing a direct link between Langmuir-type adsorption and the kinetic parameters in the BCF framework.}   
\label{coverage}
\end{figure}
As shown in Fig.~\ref{coverage}(a), $1/\theta_{\mathrm{In}}^{\mathrm{tot}}$ varies linearly with $1/F_{\mathrm{In}}$ at each temperature, indicating that the total In coverage follows a Langmuir-type behavior characterized by an effective residence time $\tau_{\mathrm{In}}^{\mathrm{tot}}$. The slope of the linear relation directly yields $\tau_{\mathrm{In}}^{\mathrm{tot}}$, while the nearly constant intercept reflects a temperature-independent saturation coverage. The extracted $\tau_{\mathrm{In}}^{\mathrm{tot}}$ is plotted as a function of $1000/T$ in Fig.~\ref{coverage}(b), showing a strong temperature dependence consistent with thermally activated surface processes.

This linearized representation provides a direct link between surface coverage and the kinetic parameters in the BCF framework. In this picture, the total In coverage $\theta_{\mathrm{In}}^{\mathrm{tot}}$ is governed by the effective residence time $\tau_{\mathrm{In}}^{\mathrm{tot}}$, while the partitioning between adatom and droplet populations determines how this coverage contributes to incorporation.

Fitting the experimental data within the binary BCF framework provides access to intermediate kinetic quantities governing In incorporation. Representative growth conditions and parameters are summarized in Table~\ref{xIn_fit_table_thiswork}; the complete dataset and fitting procedure are provided in Section V and Table S3 of the Supplementary Material.\cite{2026_Lin_APL_supplemental}  

In the following, we focus on three key quantities that characterize the In incorporation process: the In adatom coverage $\theta_{\mathrm{In}}$, the effective lower-step attachment coefficient $\kappa^\mathrm{eff}_{+,\mathrm{In}}$, and the relative In attachment efficiency $R_{\mathrm{In}}(T)$, normalized to its value at 1005~K. Here, $\kappa^\mathrm{eff}_{+,\mathrm{In}}$ depends on $\theta_{\mathrm{Ga}}$ and the adatom--droplet partitioning of surface In, whereas $R_{\mathrm{In}}(T)$ captures its temperature dependence. These quantities respectively describe the available surface In population, incorporation kinetics, and temperature-dependent attachment efficiency,  providing a kinetic perspective on how precursor fluxes and growth temperature regulate In incorporation, beyond their effect on the final alloy composition.

The dependence of $x_{\mathrm{In}}$ and $G$ on precursor flows is shown in Fig.~\ref{fit_1005}(a) and~\ref{fit_1005}(b). With increasing $F_{\mathrm{In}}$, $x_{\mathrm{In}}$ exhibits a pronounced maximum, while $G$ varies only weakly. This non-monotonic behavior reflects a redistribution of In between adatom and droplet phases: as the total coverage approaches saturation, excess In is transferred into droplets [Fig.~\ref{fit_1005}(c)], reducing the adatom population that governs incorporation. 

This transition is directly reflected in the evolution of both the In adatom coverage
$\theta_{\mathrm{In}}$
and the effective attachment coefficient
$\kappa_{+,\rm In}^{\rm eff}$
(Table~\ref{xIn_fit_table_thiswork}). Below the droplet threshold (Conds.~1 and 3), $\theta_{\mathrm{In}}$  increases rapidly with $F_{\mathrm{In}}$, indicating a larger population of mobile In adatoms available for incorporation. Above the threshold (Conds.~4 and 7), $\kappa_{+,\rm In}^{\rm eff}$ decreases as excess In becomes partitioned into droplets. With further droplet growth, the increase in
$\theta_{\mathrm{In}}$
is suppressed and eventually reversed. As a result, the In incorporation rate decreases at high $F_{\mathrm{In}}$, leading to a reduction in $x_{\mathrm{In}}$, while the Ga contribution remains nearly unchanged, resulting in only a weak variation in $G$.

In contrast, increasing the TEGa flux enhances both $x_{\mathrm{In}}$ and $G$ [Fig.~\ref{fit_1005}(d) and~\ref{fit_1005}(e)], with $G$ exhibiting an approximately linear dependence on $F_{\mathrm{Ga}}$. This behavior originates from the increase in Ga adatom density, which promotes both Ga incorporation and Ga-assisted In incorporation. Consistently, the In incorporation factor $\kappa_{+,\rm In}^{\rm eff}$ increases significantly with $F_{\mathrm{Ga}}$ (Conds.~7 and 8), reflecting the expansion of available incorporation pathways for In adatoms. At the same time, the droplet population is reduced [Fig.~\ref{fit_1005}(f)], leading to a shift of In from droplets to adatoms. These combined effects result in a simultaneous increase of $x_{\mathrm{In}}$ and $G$.

\begin{figure}
    \centering
    \includegraphics[width=0.85\linewidth]{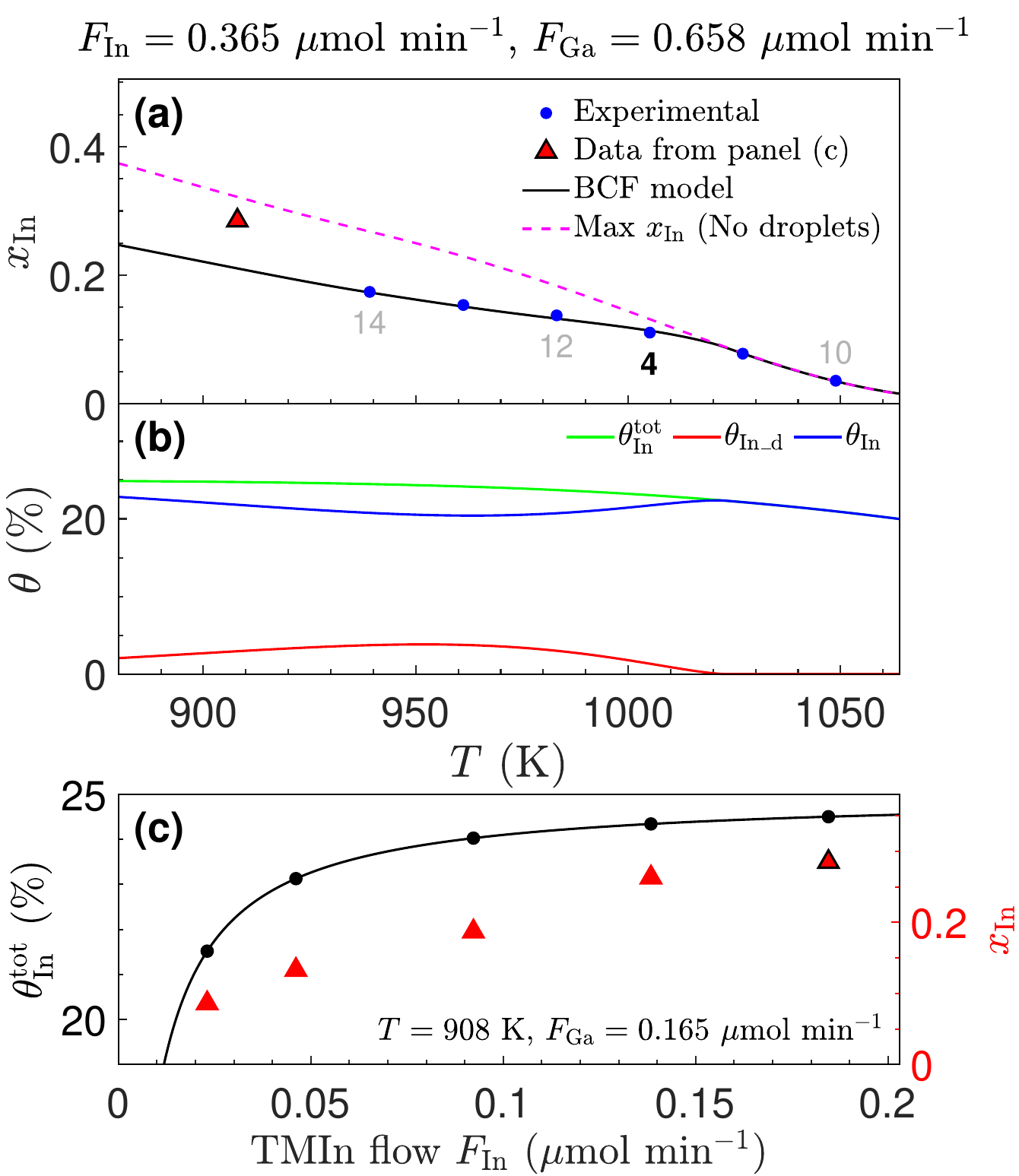}
\caption{
(a,b) Temperature dependence of $x_{\mathrm{In}}$ and surface coverages at fixed $F_{\mathrm{In}}$ and $F_{\mathrm{Ga}}$. Dashed line: calculated threshold (maximum $x_{\mathrm{In}}$ without droplets); see Eq.~(S25) in the Supplementary Material.\cite{2026_Lin_APL_supplemental} Triangles: data at 908~K [panel (c)]. 
(c) Independent measurements at 908~K showing $x_{\mathrm{In}}$ (right axis) and $\theta_{\mathrm{In}}^{\mathrm{tot}}$ (left axis) vs $F_{\mathrm{In}}$. Conds.~4, 10, 12, and 14 correspond to Table~\ref{xIn_fit_table_thiswork}. Cond.~4 is shared with Fig.~3(a).
}
\label{xInG_T}
\end{figure}

The temperature dependence of $x_{\mathrm{In}}$ [Fig.~\ref{xInG_T}(a)] further reveals the interplay between adsorption, phase partitioning, and incorporation kinetics. Upon cooling, $x_{\mathrm{In}}$ first increases rapidly, then exhibits a reduced slope, and finally increases again at lower temperatures. This behavior is directly linked to the evolution of surface coverage [Fig.~\ref{xInG_T}(b)]: decreasing temperature increases the total In coverage $\theta_{\mathrm{In}}^{\mathrm{tot}}$ due to the longer surface residence time [see Eq.~(\ref{theta_linear})]. At higher temperatures (Cond.~10), no droplets are observed. At intermediate temperature (Cond.~4), droplet formation becomes significant, reducing the adatom fraction and limiting further increase in $x_{\mathrm{In}}$. At lower temperature (Cond.~14), although $\theta_{\mathrm{In}}^{\mathrm{tot}}$ is nearly saturated, the incorporation kinetics continue to evolve, leading to a reduced droplet fraction and a renewed increase in $x_{\mathrm{In}}$. This trend is supported by the increase in both $R_{\mathrm{In}}$ and $\kappa_{+,\rm In}^{\rm eff}$ with decreasing temperature, indicating enhanced In attachment and a more efficient supply of incorporable In to the step edge.

\begin{table}[t]
\centering
\caption{
Selected representative experimental datasets used for fitting the binary BCF model, together with the corresponding fitted kinetic parameters. The data are grouped according to the parameter being varied. Here, $\kappa_{+,\rm In}^{\rm eff}$ denotes the effective attachment coefficient for In incorporation, including the effects of Ga-assisted incorporation and In partitioning into droplets.
}
\label{xIn_fit_table_thiswork}

\footnotesize
\setlength{\tabcolsep}{3pt}
\renewcommand{\arraystretch}{1.08}

\resizebox{\columnwidth}{!}{%
\begin{tabular}{cccccccccc}
\hline
\multicolumn{6}{c}{\textit{Experimental datasets}} &
\multicolumn{4}{c}{\textit{BCF model}} \\
\hline
Cond. & $T$ & $F_{\mathrm{Ga}}$ & $F_{\mathrm{In}}$ & $x_{\mathrm{In}}$
& $\theta^{\mathrm{tot}}_{\mathrm{In}}$
& $\theta_{\rm In}$ & $\kappa_{+,\rm In}^{\rm eff}$
& $R_{\mathrm{In}}$ & $\theta^{\mathrm{crit}}_{\mathrm{In}}$ \\
 & (K) & \multicolumn{2}{c}{($\mu\mathrm{mol\,min^{-1}}$)}
& (\%) & (\%)
& (\%) & ($10^{-9}\mathrm{m\,s^{-1}}$) & -- & (\%) \\
\hline
\multicolumn{10}{l}{\textit{Constant $F_{\mathrm{Ga}}$ series}} \\
\hline
1  & 1005 & 0.658 & 0.0457 & 4.6  & 14.9 & 14.9 & 2.47 & 1.00 & 22.9\\
3  & 1005 & 0.658 & 0.183  & 10.6 & 21.4 & 21.4  & 2.47 & 1.00 & 22.9\\
4  & 1005 & 0.658 & 0.365  & 10.9 & 23.1 & 22.5  & 2.41 & 1.00 & 22.9\\
\hline
\multicolumn{10}{l}{\textit{Constant $F_{\mathrm{In}}$ series}} \\
\hline
7  & 1005 & 0.658 & 0.914 & 8.1 & 24.2 & 20.9  & 2.13 & 1.00 & 22.9\\
8  & 1005 & 1.32  & 0.914 & 9.1 & 24.2 & 21.5  & 4.40 & 1.00 & 23.5\\
9  & 1005 & 0.329 & 0.914 & 6.7 & 24.2 & 20.0 & 1.02 & 1.00 & 17.2\\
\hline
\multicolumn{10}{l}{\textit{Temperature series}} \\
\hline
10 & 1049 & 0.658 & 0.365 & 3.6  & 21.0 & 21.0 & 0.896 & 0.469 & 21.0\\
12 & 983  & 0.658 & 0.365 & 13.8 & 23.7 & 20.7  & 3.65 & 1.48  & 23.6\\
14 & 939  & 0.658 & 0.365 & 17.4 & 24.5 & 20.8  & 5.54 & 2.47  & 24.1\\
\hline
\end{tabular}%
}

\end{table}

These results show that $x_{\mathrm{In}}$ is governed by coupled adsorption, phase partitioning, and incorporation kinetics rather than by a single thermodynamic parameter. Although spinodal/binodal decomposition has been proposed as a possible origin of compositional limitation in InGaN,\cite{koukitu1998thermodynamic,ho1996solid} the present CTR measurements reveal neither phase separation nor significant compositional inhomogeneity. The observed limitation of indium incorporation is therefore attributed primarily to growth-front incorporation kinetics. In a related study by Jiang \textit{et al}.,\cite{2006_Jiang_APL89_161915} metallic In droplets were found to form at an In coverage of $\sim$0.25 ML (25$\%$) when only TMIn was supplied, corresponding to a limiting case in which droplet formation is governed solely by adatom accumulation in the absence of incorporation pathways.

In contrast, during InGaN growth, the effective droplet formation threshold $\theta_{\mathrm{In}}^{\mathrm{crit}}$ is growth-condition dependent and represents the maximum surface In coverage that can be sustained by the moving growth front through incorporation.\cite{2026_Lin_APL_supplemental} Droplet formation is predicted when $\theta_{\mathrm{In}}^{\mathrm{tot}} > \theta_{\mathrm{In}}^{\mathrm{crit}}$, defining a kinetic condensation boundary.

The position of this boundary is controlled by both Ga adatom coverage and temperature. Increasing $F_{\mathrm{Ga}}$ enhances the density of available incorporation pathways, raising $\theta_{\mathrm{In}}^{\mathrm{crit}}$, while decreasing temperature increases the effective attachment efficiency of In adatoms (as reflected by $R_{\mathrm{In}}$), further shifting the boundary to higher coverage. Consequently, the kinetic condensation boundary is not tied to a fixed coverage value: it can be reached before $\theta_{\mathrm{In}}^{\mathrm{tot}}$ approaches 25\%. This demonstrates that droplet formation is governed by kinetic imbalance rather than by a universal coverage threshold.

Within this framework, the maximum attainable In composition is not governed by equilibrium thermodynamics, but is determined by the kinetic condensation boundary, where the system operates at $\theta_{\mathrm{In}}^{\mathrm{tot}} \lesssim \theta_{\mathrm{In}}^{\mathrm{crit}}$, immediately before excess In is partitioned into droplets. This boundary, represented by the dashed line in Fig.~\ref{xInG_T}(a), provides a quantitative prediction of the temperature-dependent incorporation limit.

To validate this prediction, an additional growth experiment was performed at 908~K, in which the TMIn flow was increased stepwise on a single sample. This approach minimizes droplet formation and enables access to the intrinsic incorporation limit. The growth sequence is shown in Fig.~\ref{xInG_T}(c), with details in Table~S1 of the Supplementary Material.\cite{2026_Lin_APL_supplemental} The measured 908~K data follow the predicted trend and are consistent with the calculated kinetic condensation boundary [Fig.~\ref{xInG_T}(a)]. The slight deviation from the predicted maximum attainable In composition boundary is consistent with the reduced Ga flux required under strain-limited growth conditions, which is expected to decrease the effective $\theta_{\mathrm{In}}^{\mathrm{crit}}$ according to the present model. This agreement validates the predictive capability of the present framework and establishes a physics-based strategy for identifying the optimal growth window, enabling the maximum attainable In composition to be achieved through rational selection of precursor fluxes and growth temperature rather than empirical trial-and-error.

The saturation coverage of $\sim$0.25 ML\cite{2006_Jiang_APL89_161915} is intriguingly similar to the available-site fraction of the GaN(0001)-3H(T1) reconstruction,\cite{2021_Ju_NatCommun12_1721} although the origin of this correspondence remains unclear. Such a connection, if confirmed, may suggest surface-reconstruction engineering as a route to higher In incorporation.

In summary, we combine \textit{in situ} synchrotron X-ray CTR measurements with a binary BCF model to quantify Ga--In coupled incorporation during InGaN MOVPE. The model captures the nonlinear dependence of $x_{\mathrm{In}}$ on flux and temperature arising from phase partitioning and coupled Ga--In incorporation kinetics.  The maximum In composition is governed by a kinetic balance between In adatom supply and incorporation capacity, which defines a kinetic condensation boundary that shifts with both temperature and Ga flux. The predicted temperature-dependent maximum In composition is consistent with an independent low-temperature growth experiment, establishing a quantitative framework for identifying optimal MOVPE growth conditions for high-In-content InGaN without droplet formation.

\section*{Supplementary Material}

The supplementary material includes additional experimental details, CTR analysis and fitting procedures, derivation of the binary BCF model, parameter fitting methods, supplementary figures and tables, and the complete fitting results supporting the conclusions of this work.

\section*{Acknowledgments}
This work was supported by the National Key Research and Development Program of China (Grant No. 2023YFE0124600) and the National Natural Science Foundation of China (Grant No. 62574008). The author gratefully acknowledges G. Brian Stephenson, Carol Thompson, and Jeffrey A. Eastman of Argonne National Laboratory for developing \textit{in situ} X-ray measurement techniques that greatly benefited this work.

\section*{Data Availability Statement}

The data that supports the findings of this study are available from the corresponding author upon reasonable request.

\bibliography{2024_InGaN}

\end{document}